\documentclass[12pt]{article}

\usepackage{graphicx}

\title{Solving the P/NP Problem\\ under Intrinsic Uncertainty}

\author
{Stefan Jaeger\\
CAS-MPG Partner Institute for Computational Biology \\
Shanghai Institutes for Biological Sciences \\
Shanghai 200031, China \\
E-mail: jaeger@picb.ac.cn
}

\date{}

\begin{document} 

\maketitle 

\begin{abstract}
Heisenberg's uncertainty principle states that it is not possible to compute both the position and momentum of an electron with absolute certainty. However, this computational limitation, which is central to quantum mechanics, has no counterpart in theoretical computer science. Here, I will show that we can distinguish between the complexity classes~P and~NP when we consider intrinsic uncertainty in our computations, and take uncertainty about whether a bit belongs to the program code or machine input into account. Given intrinsic uncertainty, every output is uncertain, and computations become meaningful only in combination with a confidence level. In particular, it is impossible to compute solutions with absolute certainty as this requires infinite run-time.
Considering intrinsic uncertainty, I will present a function that is in~NP but not in~P, and thus prove that~P is a proper subset of~NP. I will also show that all traditional hard decision problems have polynomial-time algorithms that provide solutions with confidence under uncertainty.
\end{abstract}

% ------------------------ Begin of main text ------------------------

\section{Introduction}

For pattern recognition problems, such as character or speech recognition, it is very common for classifiers to output confidence values indicating the confidence in their outputs~\cite{mySpringerInfConfChapter}. The output of a Turing machine, however, has always been considered a hard decision with no uncertainty involved. In fact, with its output being derived from a well-defined set of non-probabilistic rules, a Turing machine has no reason to output a confidence value, or so it seems. One might also argue that a Turing machine can emulate any method for computing confidence values, and is thus always able to output a confidence value together with its standard output, if needed. This argument misses the point though. Any method that produces confidence values when executed on a classical Turing machine is a hard decision in itself, and would therefore require a confidence value for its own computation; i.e., a confidence value for a confidence value. Obviously, this process of confidence generation cannot continue forever. What is needed instead is an intrinsic confidence, or rather an intrinsic uncertainty built into the Turing machine.

Intrinsic uncertainty has its proper place in modern physics, in particular in quantum mechanics, where the idea that we can locate objects exactly breaks down. According to Heisenberg's uncertainty principle, locating a particle in a small region of space makes the momentum of the particle uncertain; and conversely, measuring the momentum of a particle more precisely makes its position uncertain. In mathematical terms, Heisenberg's uncertainty principle can be stated as follows~\cite{kennard1927}:
\begin{equation}
\frac{\hbar}{2} \le \Delta x * \Delta p
\label{uncertaintyPrincipleFormula}
\end{equation}
where~$\hbar$ is a constant, the so-called Dirac's constant, and $\Delta x$ and $\Delta p$ denote the uncertainty about the exact location of the particle and the dispersion of its momentum, respectively. The uncertainty principle thus guarantees that the product of the errors in position and momentum is at least as large as a positive constant. It is therefore not possible to compute both the position and the momentum of a particle simultaneously with arbitrary precision. The more precisely we determine the position of the particle, the less we will know about its momentum. Conversely, the more we know about the momentum, the less we will know about the position of the particle. This uncertainty about a particles's position and momentum is established the moment the particle is observed, resulting in the measured values always being dispersed. Due to its significance, and its appealing simplicity, the uncertainty principle is one of the most important corner stones of physics. 

The intrinsic uncertainty of our physical world, as expressed in Heisenberg's uncertainty principle, motivates my integration of intrinsic uncertainty into the analysis of computational complexity. I begin my analysis with the simple observation that there is no clear allocation of tasks between program code and machine input, contrary to what the typically portrayed machine architectures might suggest. Instead, there is a chance that the program code is actually input, and vice versa, that input is actually code. This introduces uncertainty in the form of two possible interpretations of a bit string, namely, as program code or as machine input. Following an information-theoretical approach, I measure the uncertainty involved in this binary decision by computing the standard entropy. As I will show in the paper, the consideration of such an intrinsic uncertainty in all computations will ultimately lead to the fact that there exists a computable function outside~P but in~NP.

The structure of the paper is as follows:
After this introduction, Section~\ref{ComputationalModelSection} shortly repeats the main statements of complexity theory, as they pertain to the P~$\neq$~NP problem, and discusses the machine model used as a basis for the theoretical framework developed here. Section~\ref{UncertaintySection} introduces the fundamental principle of this work, i.e. the intrinsic uncertainty principle, which is central to the proof of P~$\neq$~NP, as developed in this paper. Section~\ref{ComputingSection} then discusses the consequences of the intrinsic uncertainty principle for computations performed on the theoretical machine model, and Section~\ref{SelfComputationSection} presents the main proof technique used to show that P~$\neq$~NP, namely the concept of self-computation. The following two sections, Section~\ref{SigmoidFunctionSection} and Section~\ref{EntropySection}, establish the connection with information theory, touching on subjects like sigmoid function, golden ratio, and, most importantly, entropy. Next, Section~\ref{ComputationalPrecisionSection} and Section~\ref{ComplexitySection} introduce three lemmas addressing problems entailed by the intrinsic uncertainty principle; in particular, computational precision, memory and code size requirements, and code complexity. At the end of the paper, the results of all three lemmas combined will lead to the conclusion that there exists a computable function that is in~NP but not in~P.

\section{Computational Model}
\label{ComputationalModelSection}

In this paper, I will confine myself to introducing the bare essentials of complexity theory; listing only the informal definitions of the major complexity classes~P and~NP. For more information on the formal definition of a Turing machine, NP-completeness, and other concepts of complexity theory, I refer readers to the relevant literature; e.g.~\cite{garey-book,cormen}.

The complexity class~P contains all problems that are solvable by a deterministic Turing machine in polynomial time, which means they can be solved in time $O(n^k)$ for some constant~$k$, where~$n$ is the size of the input to the problem. On the other hand, the complexity class~NP contains all problems that are verifiable by a deterministic Turing machine in polynomial time, which means that any given potential solution to the problem can be verified in time polynomial in the size of the input to the problem. Equivalently, the class~NP can be defined as the set of problems that are solvable by a non-deterministic Turing machine in polynomial time, where a non-deterministic Turing machine can compute many operations in parallel. While a deterministic Turing machine has at most one entry for each combination of tape symbol and state in its action table, and can thus perform only one possible operation at a specific time instant during run-time, a non-deterministic Turing machine can have multiple entries in its action table. We can think of a non-deterministic Turing machine as a lucky guesser performing a deterministic computation by always choosing, luckily, the correct operation leading to the solution.

We can directly follow that any problem in~P is also a member of~NP; i.e. P~$\subseteq$~NP, because any problem in~P can be directly solved in polynomial time without the need of verifying all potential solutions. However, the complexity class~NP is known to include many problems, called NP-complete problems, for which no polynomial-time algorithms are known. An open question in complexity theory is, therefore, whether polynomial-time algorithms actually exist for these problems. In other words, the open question is whether or not~P is a proper subset of~NP. This is the well-known P~$\neq$~NP problem.

This paper will largely abstract from the machine model doing the actual computation on a set of instructions and given input. Let us merely assume that the machine contains a tape of binary storage cells as its only memory. The tape can have infinite length, like the tape of a Turing machine. In our case, however, the tape will contain both the program code to be executed and its input. Accordingly, a certain percentage of the memory used by the machine will be consumed by the coding of the program instructions, e.g. the coded transition table in case of a Turing machine, while the remaining memory cells serve as memory. When compared to the standard Turing model, which stores the program code in a separate ROM module, this seems like a minor difference because it does not affect the computational expressiveness. Yet, it puts code and memory on equal terms; paving the way for the approach followed here. In fact, the artificial separation made by the traditional Turing machine between program code and memory goes against our practical programming experience. It is often possible to solve a coding task in multiple ways by substituting code with memory. We can either choose to implement very sophisticated program code on simple data structures or instead use a concise code in combination with more elaborated data structures. Both implementations will solve the coding problem, and a programmer's preference will usually depend more on software engineering principles and less on theoretical aspects. While these thoughts are perhaps more of a philosophical exercise, the fact that today's complexity theory closes the eyes to code complexity is of severe theoretical significance. In particular, the definitions of the major complexity classes NP and~P contain no reference to the size of the program code. They concentrate exclusively on the run-time of the program measured with respect to the input size~\cite{cormen}. This is not to say that these definitions are wrong, but they do not describe the core of problem complexity adequately. For instance, by implementing a look-up table directly in the program code that contains the correct output for all or part of the program's possible input values, much like a case-statement lists the output alternatives for each possible input in a procedural programming language, we can easily speed up the run-time of a program at the cost of increasing the size of its code. Of course, classic theory assumes that we do not know such a program in advance and that we have to compute the individual solutions first before we can implement the program, in which case the traditional complexity theory takes effect. Nevertheless, this example shows that program size plays a role and should be considered when discussing the complexity of programs.

This being said, let us assume in the following that we have a Turing-complete machine architecture that can access a binary storage tape of infinite length. Like in the classical decision problem scenario, our machine will be given an input, written on its tape, that it either accepts or rejects. In order to perform this computation, the machine requires a certain number of memory cells, from which one part acts as program code and the other serves as memory accommodating the input. The program code represents the known facts, while the input is random in the sense that it must be accepted or rejected by the machine based on the known facts and relationships coded in the program. Note that program code and memory do not need to occupy coherent blocks in the bit string. Both can be interweaved and distributed all across the bit string. Furthermore, the model provides for self-modifying code, as a program can access and modify its own code.

\section{Uncertainty}
\label{UncertaintySection}

Motivated by Heisenberg's uncertainty principle, which states that there are conflicting quantities that we can compute with infinite precision only individually but not simultaneously, this section is going to introduce a logical uncertainty principle, namely the intrinsic uncertainty principle. It basically says that there is no absolute truth and that it is impossible to solve problems individually; they can only be solved in pairs. Based on this principle, the following sections are developing a function that is in~NP but not in~P.

Given a partition of $N$~bits on our Turing machine's tape into program code and input, the intrinsic uncertainty principle states that we do not know what part is program code and which part is input. Accordingly, there are two possible interpretations: We can use the first part as program code and the second one as the input that needs to be verified, or the other way around, use the first part as input and the second one as program code. Since we do not know which interpretation is correct, our only choice is to perform both interpretations simultaneously and accept both of their results, if they produce any, as correct. Nevertheless, we do not have to accept them on equal terms. We can accept the interpretation whose program code encompasses the larger number of bits as more likely. Then, the relative size of the assumed program code; i.e. the percentage of program bits in the bitstring, can serve as a confidence measure. Note that we face this ambiguity for any partition of the bitstring into program code and memory. In the extreme case that all the bits are considered program code, there is no memory and therefore nothing to compute, and thus no uncertainty. Similarly, if we consider all the bits to be random memory, then there is no program code and again nothing to compute, though this time the uncertainty is maximum as the bitstring is completely random. The original motivation lying in Heisenberg's uncertainty principle shows here. When the uncertainty for one interpretation is minimum; i.e. its program code takes up the entire bitstring, the uncertainty for the opposite interpretation becomes maximum as it considers the entire bitstring as random memory. Conversely, if the uncertainty for one interpretation is maximum, the opposite interpretation's program code consists of the entire bitstring with no place for random bits or uncertainty. Analogous to Heisenberg's uncertainty principle, it is not possible to compute both interpretations simultaneously with infinite precision; i.e. no uncertainty.

\section{Computing}
\label{ComputingSection}

Let~$S$ be an initial string partitioned into two subsets~$S_{1}$ and~$S_{2}$ on the tape of our Turing machine. Under the intrinsic uncertainty principle, we have to perform two computations simultaneously; one for each of the two possible interpretations of the two subsets. This will produce two different results, each associated with a confidence value that is proportional to the program code generating it; i.e. the size of the subset serving as program code. In the following, let us assume that at least one of the two subsets~$S_{1}$ and~$S_{2}$ codes a member function~$f$ of~NP. Then, one of our two computations is the actual implementation of~$f$ and will produce its output by using one of the two subsets as input. The other computation, however, uses the input to~$f$ as program code and the code as input. There are no constraints regarding the complexity of this second computation. It neither needs to stop nor does it need to implement a function of~NP. Nevertheless, whatever the behavior of the second computation, it produces a result that stands in competition with the result of the first computation. 

In addition to its tape, our Turing machine consists of logic that performs the actual computation of the initial string's interpretations. The storage for this logic needs to be added to the program code of any interpretation of the initial string. This is an important detail that is mostly ignored in traditional complexity theory, which abstracts the underlying hardware from the logical program execution. The present paper, however, will explicitly use the functionality of this hardware. In fact, let us assume that there is an ``outside'' program~$T$ that implements our Turing machine by reading the initial string and executing its interpretations. We do know nothing about~$T$ other than its mere existence. In particular, we know nothing about its size. Nonetheless, like in traditional complexity theory, let us assume that it does not add to the complexity of the program coded in~$S_{1}$ or~$S_{2}$.

Furthermore, let us assume that for each terminating computation, $T$~returns a confidence value proportional to the program code executed. In fact, under the intrinsic uncertainty principle, a result only becomes meaningful when associated with a confidence value, as there is always the possibility that the implemented interpretation is wrong. We also need to take into account that the confidence value returned by~$T$ is the result of a probabilistic decision between two interpretations, and as such is uncertain as well. It may therefore be necessary to compute a confidence for the confidence value itself in order to reach a higher precision. However, in order to compute a confidence value for the output of~$T$, additional logic is needed. The new logic will extend the already existing logic in~$S$, thus leading to an ultimately higher confidence value. Mathematically, the higher confidence value can be computed by applying~$T$ to itself, and thus by $T(T(S))$. This recursive function call will then produce a higher confidence value at the cost of a growing~$S$, given that $T(S)$ is a string on the Turing machine's tape; with $|S| < |T(S)|$, where~$|S|$ and $|T(S)|$ denote the relative number of bits in~$S$ and~$T(S)$, respectively. The following sections will elaborate on the theoretical implications of this recursive self-computation and its connection with the P~$\neq$~NP problem.

\section{Self-Computation}
\label{SelfComputationSection}

Let $T^{2}(S) = T(T(S))$ be a self-computation of~$T$ on~$S$. This means that $T$ emulates its own behavior by running our Turing machine with the initial string~$T(S)$, which contains both the program code for~$T$ and the input string~$S$. The latter, in turn, contains the input to be processed by~$T$ with its two subparts~$S_{1}$ and~$S_{2}$. Thus, $T$ operates on a copy~$T$ of itself that is given an input~$S$, computing the function $T(T(S))$ in the process.

Under the intrinsic uncertainty principle, we do not know whether~$T$ or $T(S)$ is the actual code to execute. Nevertheless, let us assume that~$p$ is the correct confidence value denoting the relative size of the program code, with $0 \le p \le 1$. Then, given the facts that $|T|+|T(S)|=1$ and $|T| \le |T(S)|$, we need to distinguish between three different cases:
\begin{itemize}
\item{$\bf p < 0.5$:}
In this case, the smaller part~$T$ of $T(T(S))$ acts as program code.
\item{$\bf p > 0.5$:}
Here, the larger part~$T(S)$ of $T(T(S))$ is the program code.
\item{$\bf p = 0.5$:}
When both parts $T$ and $T(S)$ have the same size, then both interpretations are possible.
\end{itemize}

Note that we cannot compute $T(T(S))$, and thus~$p$, directly without adding further logic because we always need the functionality of one~$T$ for the actual implementation of our Turing machine. So, with limited resources, we have no choice other than modeling $T(T(S))$ with $T(S)$ as the initial string on the tape of our Turing machine. It follows that any confidence measurement~$o$ of our Turing machine will be observed with respect to~$|T(S)|$. In particular, the Turing machine will always set the smaller part of $T(T(S))$; namely $T$, into relation to the larger part~$T(S)$, measuring either the program code or the input. Again, we need to distinguish three different cases of~$p$:
\begin{itemize}
\item{$\bf p < 0.5$:}
Here, $T$ serves as the program code in $T(T(S))$, with $|T| = p$ and $|T(S)| = 1-p$. The Turing machine thus measures $o = \frac{p}{1-p}$.
\item{$\bf p > 0.5$:}
For~$p$ larger than~$0.5$, $T(S)$ serves as the program code in $T(T(S))$, with $|T(S)| = p$ and $|T| = 1-p$. The Turing machine will therefore measure $o = \frac{1-p}{p}$.
\item{$\bf p = 0.5$:}
With $|T| = |T(S)|$ both interpretations are possible; i.e. both~$T$ and~$T(S)$ can act as program code or input, respectively, and the Turing machine measures $o=1$.
\end{itemize}

\section{Sigmoid Function}
\label{SigmoidFunctionSection}

Knowledge of the observation~$o$ allows a Turing machine to compute the different possibilities for the true confidence~$p$ by distinguishing between the different cases just mentioned in the previous section, and simply resolving~$o$ for~$p$. Here, instead of resolving~$o$ directly for each~$p$, I take a detour by first introducing the expected information~$K$ for each~$p$, and then resolving~$K$ for~$p$. This will lead to the same essential result, but has the additional advantage of introducing important concepts, most notably the sigmoid function. We obtain the expected uncertainty~$K$ for the observation~$o$ by multiplying its logarithm with the true probability~$p$ as follows:
\begin{eqnarray}
\label{expectedInformationFormula}
K & = & -p * \ln\left( o \right) \\
\Longleftrightarrow \quad o & = & e^{-\frac{K}{p}}
\label{observationFormula}
\end{eqnarray}
A direct implication of Eq.~\ref{expectedInformationFormula} is that the observation~$o$ becomes the residual part of an exponential distribution~$D$ with expectation value~$p$ as in Eq.~\ref{observationFormula}, with $D = 1-e^{-\frac{K}{p}}$. While I am not going to explore this further, I should mention that the relationship with the exponential distribution leads to an information fusion method that can be applied to machine learning and pattern recognition problems, like for instance classifier combination~\cite{mySpringerInfConfChapter,jaegerLAMPReport,jaegerLAMPReportII}.

Let us now insert the different possibilities for~$o$ into Eq.~\ref{expectedInformationFormula}. For $p \le 0.5$, we obtain the following~$K$:
\begin{equation}
K = -p * \ln\left(\frac{p}{1-p} \right)
\label{expectedInformationFormulaI}
\end{equation}
Resolving Eq.~\ref{expectedInformationFormulaI} for~$p$ then provides the first possible value for~$p$:
%\begin{equation}
%p = \frac{1}{1 + \frac{1}{e^{-\frac{K}{p}}}}
%\label{pFormulaI}
%\end{equation}
\begin{equation}
p = \frac{1}{1 + e^{\frac{K}{p}}}
\label{sigmoidFormulaI}
\end{equation}
Using the relationship in Eq.~\ref{observationFormula}, this result can be further simplified to the following form:
\begin{equation}
p = \frac{1}{1 + \frac{1}{o}}
\label{pFormulaI}
\end{equation}

Analogously, we can perform the same computation steps for the observation~$o$ under $p \ge 0.5$, which leads to the following results and the second possible value for~$p$:
\begin{eqnarray}
\label{expectedInformationFormulaII}
K & = & -p * \ln\left(\frac{1-p}{p} \right) \\
\label{sigmoidFormulaII}
\Longleftrightarrow \quad p & = & \frac{1}{1 + e^{-\frac{K}{p}}} \\
\label{pFormulaII}
\Longleftrightarrow \quad p & = & \frac{1}{1 + o}
\end{eqnarray}
Note that the right-hand side of Eq.~\ref{sigmoidFormulaII} is the well-known sigmoid function, which plays an important role in signal processing for both natural as well as artificial neural networks~\cite{hodgkin1990,bishopbook,hechtNielsenBook}. For a given observation~$o$, Eq.~\ref{sigmoidFormulaI} and Eq.~\ref{sigmoidFormulaII} thus show that the two possible values of~$p$ are the output of a sigmoid function with input~$K$ or~$-K$, respectively.    

Another result following from Eq.~\ref{expectedInformationFormulaII} is that the measured value~$o$ equals the true probability~$p$ when~$o$ is equivalent to the Golden Ratio. The following equation expresses this relationship:
\begin{eqnarray}
p & = & \frac{1-p}{p} \label{goldenRatioFormula} \\
\Longleftrightarrow \quad p & = & 0.618... \quad (\mbox{or}\;\; p = -1.618...) \nonumber
\end{eqnarray}
There are precisely two possible values satisfying Eq.~\ref{goldenRatioFormula}, namely $\varphi \approx 0.618$ and $\Phi \approx -1.618$. These values define the so-called Golden Ratio, or Golden Mean as it is sometimes called. The Golden Ratio is an irrational number, or rather two numbers, describing the proportion of two quantities~\cite{livioGoldenRatioBook,huntleyGoldenRatioBook}: Two quantities are in the golden ratio to each other, if the whole is to the larger part as the larger part is to the smaller part; with the whole being the sum of both quantities.

\section{Entropy}
\label{EntropySection}

According to the framework set out above, any observation~$o$ will leave us with two possibilities for the true value of~$p$. Therefore, with two values possible for the true~$p$, any decision we make in favor of one of the two values will always be fraught with uncertainty. Nevertheless, it is possible to quantify the uncertainty involved by computing the entropy of the decision process~\cite{shannon}. We can do so by making use of the fact that both possible values for~$p$ in Eq.~\ref{pFormulaI} and Eq.~\ref{pFormulaII}, respectively, add up to one:
\begin{equation}
\frac{1}{1+\frac{1}{o}} + \frac{1}{1+o} = 1
\label{sumOneFormula}
\end{equation}
This allows us to obtain the entropy~$e$ in the standard way, independent from the actual~$p$ computed for a given~$o$:
\begin{equation}
e = -p * \ln(p) - (1-p) * \ln(1-p)
\label{entropyFormula}
\end{equation}
The entropy~$e$ thus describes the uncertainty involved in the decision on a particular~$p$; i.e., the decision on whether the true~$p$ is smaller or larger than~$0.5$.

\section{Computational Precision}
\label{ComputationalPrecisionSection}

In view of the uncertainty that our Turing machine~$T$ faces when choosing one of the two possible interpretations of the input string~$S$, let us introduce a threshold value as an integrated part of~$T$ in form of a given entropy~$e'$. The threshold value then imposes a confidence level on~$T$, letting the Turing machine accept input~$S$ only if the uncertainty in its interpretation is less than~$e'$. In other words, the threshold determines the precision of the output produced by~$T$, and thus the precision of~$T$ itself. For instance, in the extreme case of the threshold being equal to zero; i.e. $e'=0$, we do not permit any uncertainty in the decision process and only allow $T$ to output a result if it is absolutely certain. Consequently, for a self-computation $T(T(S))$, the only acceptable measurement~$o$ under this restriction is~$o=0$, in which case both~$T$ and~$S$ together contain either exclusively program code or input, and~$T$ has nothing to compute at all. On the other hand, if we permit all decisions independent of their uncertainty, including decisions involving maximum entropy, which means $e'=1$ when measuring information with the base~$2$ logarithm, then~$o$ can be as high as~$1$ and~$T$ may be similar in size to~$S$, with $|T|=|S|=0.5$ in the extreme case of $e'=1$.  

The following Lemma states an important result on the precision we can achieve for an arbitrary input string~$S$. Together with two other lemmas that the following section is going to introduce, it will lead to the conclusion that there exists a member function of~NP that is not in~P.\bigskip

{\noindent\bf Lemma~$1$:} For every input string~$S$ partitioned into program code~$S_{1}$ and data input~$S_{2}$, or vice versa, there exists a Turing machine~$T$ that simulates~$S_{1}$ on~$S_{2}$ with arbitrary precision.\bigskip

\noindent Proof: Let $S$ be a string partitioned into program code and data input. Furthermore, let~$T_{0}$ be a Turing machine that simulates the program code of~$S$ on the data part of~$S$. Obviously, such a Turing machine~$T_{0}$ does exist. In order to show that a Turing machine can perform the simulation with arbitrary precision, it suffices to prove that, for any given lower bound~$c$ on the entropy, there exists a Turing machine~$T$ that can do the simulation with entropy~$e<c$. If the entropy computed by~$T_{0}$ is smaller than~$c$, then we have already found such a Turing machine; i.e. $T=T_{0}$. Otherwise, we can gradually reduce the entropy of~$T_{0}$ by iteratively simulating the computation of~$T_{0}$ with~$T_{0}$ itself; i.e. by computing $T_{0}^{k}(S)$ with $k>1$. For instance, if~$T_{0}$ measures $o=|S_{1}|=\frac{1}{3}$, then, according to Eq.~\ref{pFormulaI} and Eq.~\ref{pFormulaII}, there are two different possible interpretations for $T_{0}(T_{0}(S))$, namely the first interpretation based on $p=\frac{1}{4}$ and the second interpretation based on $p=\frac{3}{4}$. Selection of the smaller value; i.e. $p=\frac{1}{4}$, as the new observation~$o$ produces another two possible interpretations for $T_{0}^{3}(S)$ based on $p=\frac{1}{5}$ and $p=\frac{4}{5}$, respectively. Note that the entropy of the decision for $T_{0}^{3}(S)$ is smaller than the entropy of the previous decision for $T_{0}^{2}(S)$. If we continue this process iteratively by always choosing, among the two possible interpretations for $T_{0}^{k}(S)$, the smaller value for~$p$ as observation for $T_{0}^{k+1}(S)$, we can reduce the entropy gradually until we fall below the given threshold~$c$. According to Eq.~\ref{pFormulaI}, the following continued fraction describes the series of smaller values for~$p$, resulting in a monotonously decreasing entropy:
\begin{equation}
p = \frac{1}{1+\frac{1}{1+\frac{1}{1+...}}}
\label{fractionFormula}
\end{equation}
Once we have found a~$k_{0}$ for which the entropy~$e$ of $T_{0}^{k_{0}}(S)$ is smaller than~$c$, we can make a final decision on an interpretation that will produce the same result as $T_{0}(S)$. The uncertainty in our decision making process is then less than~$c$ and our decision will determine the entire computation. Hence, we have found a Turing machine~$T$ with $T(S)=T_{0}^{k_{0}}(S)=T_{0}(S)$ and~$e<c$. This means~$T$ computes the same result on~$S$ as $T_{0}$ does on~$S$, but with an entropy~$e$ lower than~$c$; albeit with $e>0$.\quad qed\bigskip

\section{Problem Complexity}
\label{ComplexitySection}

This section will introduce two more lemmas that are necessary to show the final conclusion. Lemma~$2$ first makes a statement regarding the size of the code and memory required by a Turing machine~$T$ in order to do a simulation, before Lemma~$3$ establishes a connection to the complexity class~NP.\bigskip

%{\noindent\bf Lemma~$2$:} Let~$S$ be a random input string partitioned into %program code~$S_{1}$ and data~$S_{2}$, or vice versa. Every Turing machine~$T$ %that simulates the program code of~$S$ with arbitrary precision needs either %infinite memory or infinite code.\bigskip

{\noindent\bf Lemma~$2$:} Every Turing machine~$T$ that simulates the program code of any input~$S$ with arbitrary precision, where~$S$ is partitioned into code~$S_{1}$ and data~$S_{2}$ or vice versa, needs either infinite memory or infinite code.\bigskip

\noindent Proof: Let~$T$ be a Turing machine doing the simulation on any input~$S$, using~$S_{1}$ or~$S_{2}$ as program code. In order to show that~$T$ needs either infinite memory or infinite code, we can proceed analogously to the proof of Lemma~$1$ by proving that either the memory size or code size can exceed any given upper bound~$c$. We can again use the concept of self-computation to do so: Let $S'=T(S)$ be an input string describing the application of~$T$ to input~$S$. Then, we apply Eq.~\ref{pFormulaI} to the self-computation $T(S')=T(T(S))$, with the result that for sufficiently low entropy values either the memory size or code size would need to exceed~$c$. In particular, there exists a lower bound~$e'$ on the entropy so that all decisions with entropy~$e<e'$ need memory or code larger than~$c$, depending on what observation~$o$ actually measures; i.e. program code or memory. We can specify the observation~$o'$ leading to~$e'$ as follows: 
\begin{equation}
o' = \min\left(\frac{L(T)}{c},\frac{L(S)}{c} \right)
\label{boundaryObservationFormula}
\end{equation}
with~$L(x)$ denoting the length of~$x$, measured in number of bits, and $0 < L(T),L(S) < c$. According to Eq.~\ref{pFormulaI}, the smaller value~$p'$ of the two possible values of the true probability is 
\begin{equation}
p' = \frac{1}{1 + \frac{1}{o'}}
\label{pFormulaIBound}
\end{equation}
From~$p'$, we can directly compute the lower bound~$e'$ as
\begin{equation}
e' = -p' * \ln(p') - (1-p') * \ln(1-p')
\label{entropyBoundFormula}
\end{equation}
The Turing machine~$T$ will then breach the upper bound~$c$ for any decision it makes with an uncertainty less than~$e'$.\quad qed\bigskip

Lemma~$3$ is now going to show a relationship between a Turing machine with a pre-specified upper entropy bound~$e'>0$ and the complexity class~NP.\bigskip

{\noindent\bf Lemma~$3$:} Let~$T_{e'}(S)$ be a Turing machine that simulates the program code of input~$S$ with precision~$0<e<e'$, where each input string~$S$ is partitioned into code~$S_{1}$ and data~$S_{2}$ or vice versa. Then, $T_{e'}(S)$ is in~NP if $S_{1}$ or~$S_{2}$ codes a decision function in~NP.\bigskip

%\noindent Proof: First, $\Rightarrow$: Let $T_{e'}(S)$ be a Turing machine %computing a decision problem in~NP based on input~$S$. Then, $T_{e'}(S)$ %simulates one of the two possible decision problems in~$S$ and returns a %decision based on a program coded by either~$S_{1}$ or~$S_{2}$. With %$T_{e'}(S)$ computing a decision problem in~NP, it follows that at least one %of the two substrings $S_{1}$ and~$S_{2}$ needs to code a decision problem %in~NP, because otherwise the decision problem computed by $T_{e'}(S)$ would %fall out of~NP, which would be a contradiction.
%Second, $\Leftarrow$:

\noindent Proof: Let~$S$ be an input string with either $S_{1}$ or~$S_{2}$ coding a decision function in~NP. Then, $T_{e'}(S)$ can be implemented as a Turing machine that simulates both possible interpretations in parallel. With one of the two simulations being the execution of an actual member function of~NP, we are guaranteed to obtain a decision in polynomial time on a non-deterministic Turing machine. Moreover, we can expect a decision independently of the behavior of the second simulation, which does not necessarily need to return a result. It follows that~$T_{e'}(S)$ computes a decision problem in~NP with uncertainty~$e$. According to the proof of Lemma~$1$, the precision of~$T_{e'}$ can be gradually increased by the concept of self-computation until the entropy~$e$ falls below any given threshold~$e'$. In particular, there exists a $k \ge 1$ for which the entropy of $T_{e'}^{k}(S)$ falls below~$e'$, with $T_{e'}^{k}(S) = T_{e'}(S)$. Furthermore, with~$T_{e'}(S)$ being a member function of~NP, the concatenation $T_{e'}^{2}(S)$ of~$T_{e'}(S)$ is also in~NP. Arguing inductively, it follows that $T_{e'}^{k}(S)$ is in~NP, too. In the form of~$T_{e'}^{k}(S)$ we have thus found an implementation of~$T_{e'}(S)$ that is both in~NP and simulates the program code of~$S$ with precision~$e<e'$.\quad qed

\section{Conclusion}

With the three proved lemmas above it is now possible to show that, under the intrinsic uncertainty principle, there exists a member function of~NP that is not in~P. The idea is to ``blow-up'' the precision of a member function of~NP so that it is guaranteed to fall out of~P. In particular, let $T_{e'}^{\mbox{NP}}(S)$ be a Turing machine that simulates a specific problem of~NP on any given input with a precision not larger than~$e'$, and with both problem and input being coded as substrings~$S_{1}$ and~$S_{2}$ of~$S$. Then, according to Lemma~$1$, $T_{e'}^{\mbox{NP}}(S)$ is always a computable function independent of the $e'$~chosen, as we can perform the simulation with arbitrary precision, and in particular with a precision smaller than~$e'$. According to Lemma~$3$, it also follows that $T_{e'}^{\mbox{NP}}(S)$ is a member function of~NP because either $S_{1}$ or~$S_{2}$ codes a decision function in~NP.

We can specify an uncertainty threshold~$e'$ for which $T_{e'}^{\mbox{NP}}(S)$ is not in~P, using again the concept of self-computation and the threshold calculation in the proof of Lemma~$2$. For this purpose, let~$e'$ be a dynamic threshold depending on the size of the input~$S$, so that~$e'$ in fact now becomes a function $e'(L(S))$, and $T_{e'}^{\mbox{NP}}(S)$ therefore becomes $T_{e'(L(S))}^{\mbox{NP}}(S)$, where $L(S)$ denotes the length of input~$S$ measured in number of bits. Note that all three lemmas introduced above can be formulated in a way that their statements hold for dynamically changing thresholds, thus remaining valid for thresholds depending on the size of input~$S$. For the following specific threshold $e'(L(S))$, we can then conclude that $T_{e'(L(S))}^{\mbox{NP}}(S)$ cannot be in~P:
\begin{equation}
e'(L(S)) = -p'(L(S)) * \ln(p'(L(S))) - (1-p'(L(S))) * \ln(1-p'(L(S)))
\label{entropyDynamicBoundFormula}
\end{equation}
where the probability function $p'(L(S))$ is defined as
\begin{equation}
p'(L(S)) = \frac{1}{1 + 2^{L(S)}}
\label{pFormulaIDynamicBound}
\end{equation}
According to Lemma~$1$, we know that there exists a self-computation that reaches the required precision~$e'(L(S))$, with either $S_{1}$ or~$S_{2}$ containing the program code of $T_{e'}^{\mbox{NP}}$:
\begin{equation}
T_{e'(L(S))}^{\mbox{NP}}(S) = T_{e'\left(L \left(T_{e'(L(S))}^{\mbox{NP}}(S) \right) \right)}^{\mbox{NP}} \left(T_{e'(L(S))}^{\mbox{NP}}(S) \right)
\label{selfComputationFormula}
\end{equation}
According to Eq.~\ref{pFormulaI} with $p'(L(S)) = \frac{1}{1 + \frac{1}{o'}}$, we also know, given the definitions of $e'(L(S))$ and $p'(L(S))$, that this self-computation entails an interpretation implying the following observation~$o'$:
\begin{equation}
o' = \frac{1}{2^{L(S)}}
\label{dynamicBoundaryObservationFormula}
\end{equation}
This observation guarantees that the input~$S$ occupies at most one percent of the input to the self-computation, which is exponential in the size of~$S$. The Turing machine $T_{e'(L(S))}^{\mbox{NP}}(S)$ therefore requires either an exponential number of execution steps or memory cells if it actually has to perform the self-computation; with both requirements leading to an exponential run-time. Given that no implementation of $T_{e'(L(S))}^{\mbox{NP}}(S)$ can avoid input leading to a self-computation, there exists always an input that requires exponential run-time for any particular implementation. Consequently, $T_{e'(L(S))}^{\mbox{NP}}(S)$ cannot be a member of~P.

At the end of this paper, let me comment on how intrinsic uncertainty relates to the hard decisions of the traditional Turing machine. There are two possible ways to deal with hard decisions under intrinsic uncertainty: First, we can insist on hard decisions for every computation; i.e., an entropy threshold equal to zero. In this case, however, every program would have an infinite run-time. Second, we can acknowledge the fact that hard decisions in the traditional sense are not possible, and could agree on a small, but positive, entropy threshold as a confidence level below which every decision is considered hard. This more realistic option has consequences for the traditional hard decision problems, which can only be solved in pairs: It follows that a Turing machine can simply output any arbitrary answer for a traditional decision problem and, by doing so, can increase the precision of the combined decision by means of self-computation. This is possible because any additional uncertainty will reduce the observed probability~$o$, and thus also reduce the uncertainty of the combined decision according to Eq.~\ref{sumOneFormula} and Eq.~\ref{entropyFormula}. In this way, we can compute solutions to any hard decision problem with arbitrary precision~$e>0$ in polynomial time. Alas, there is no way of knowing the correct solution. The more uncertainty we allow in the output, and thus the more we increase the precision, the less we know about the true answer. This is again reminiscent of Heisenberg's uncertainty principle~\cite{jaegerLAMPReportII}. Nevertheless, it follows that all traditional hard decision problems are in~P. Note that this also includes heuristics that make explicit statements about the quality of their approximations, which are themselves hard decisions. The distinction between~P and~NP comes to light when we consider soft decision problems that include explicit entropy thresholds. 

\begin{figure}
\centering 
%\hspace{-0.5cm}
\includegraphics[width=7cm]{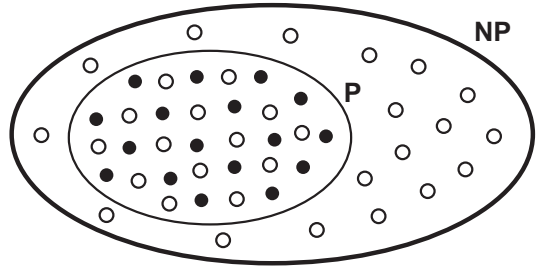}
\caption{The complexity classes~P and~NP with hard decision problems (dots) and soft decision problems (circles).} 
\label{PNPFigure} 
\end{figure}

Figure~\ref{PNPFigure} shows the hierarchy of complexity classes as it presents itself after considering intrinsic uncertainty.  
The traditional hard decision problems are represented by black dots, while the new soft decision problems are circles. According to Figure~\ref{PNPFigure}, the complexity class~P is a proper subset of class~NP. Nevertheless, all traditional hard decision problems are members of~P, including the NP-complete problems and heuristics. This means that all of the traditional decision problems have in fact polynomial-time algorithms, which may be an unexpected result. Of course, the simple method described above for generating polynomial-time algorithms will hardly satisfy programmers looking for faster algorithms in the traditional sense.

The heart of the P/NP problem lies in the distinction between hard and soft decisions. While hard decision problems can only be in~P, soft decision problems can be in both~P and~NP. This paper has shown the existence of soft decision problems that are in~NP but not in~P. Nevertheless, soft decision problems can also be in~P. For instance, using a maximum entropy threshold of~$1$, all soft decisions on traditional polynomial-time problems, are in~P.

All facts considered, this paper comes to the conclusion that~P is not equal to~NP.

%\section*{Acknowledgment}
%The support of this research by the Department of Defense under contract %MDA-9040-2C-0406 is gratefully acknowledged.

% ------------------------ End of main text ------------------------

% Bibliography

\bibliography{PNP}
\bibliographystyle{plain}

\end{document}